\documentclass[pre,twocolumn,a4paper,superscriptaddress]{revtex4}
\usepackage{amssymb}
\usepackage{wasysym}
\usepackage{graphicx}
\usepackage{epsfig}
\usepackage{subfigure}
\begin{document}

\title{When more of the same is better}

\author{Jos\'e F. Fontanari}
\affiliation{Instituto de F\'{\i}sica de S\~ao Carlos,
  Universidade de S\~ao Paulo,
  Caixa Postal 369, 13560-970 S\~ao Carlos, S\~ao Paulo, Brazil}


\begin{abstract}
Problem solving (e.g., drug design, traffic engineering, software development) by task forces represents a substantial portion  of the economy  of developed countries. Here we use  an agent-based model of cooperative problem solving systems to study the influence of  diversity on the performance of a  task force. We assume that  agents cooperate by exchanging information on their partial success and use that information to imitate the more successful agent in the system -- the model. The agents differ only in their propensities to copy  the model. We find  that,
for easy tasks,  the optimal organization is a homogeneous system composed  of agents with the highest possible copy propensities.
For difficult tasks, we find that  diversity can prevent the system from being trapped in sub-optimal solutions. However, when
the system size is adjusted  to maximize  performance  the homogeneous systems outperform the heterogeneous systems, i.e.,  for optimal performance, sameness should be preferred to diversity.

\end{abstract}

\maketitle

\section{Introduction}

Understanding the factors that influence the capability of a group of individuals  to solve problems  is a central issue  on collective intelligence \cite{Huberman_90,Clearwater_91,Fontanari_14} and on organizational design \cite{Page_07,Lazer_07,Herrmann_14}, nonetheless the meager interchange of ideas  between these two research areas. 
Conventional wisdom says that a group of cooperating individuals can solve a problem faster than the same group of individuals working in isolation, and that the higher the diversity of the group members, the better the performance.
Although   there has been some progress on the  quantitative understanding of  the factors that make cooperative group work effective \cite{Clearwater_91,Hong_04,Rendell_10},  only very recently    a workable minimal agent-based model of distributed cooperative problem solving system was proposed \cite{Fontanari_14} (see also \cite{Lazer_07}). Here we  build on that model to dispute  some common-sense views of the benefits of  diversity in group organization.

We consider a distributed cooperative problem solving system
 in which agents cooperate by broadcasting messages informing on their partial success towards the completion of the goal and use this information to imitate the more successful agent (model) in the system. In doing so, we follow Bloom in conferring imitative learning the central role
in the emergence of collective intelligence:  ``Imitative learning acts like a synapse, allowing information to leap the gap from one creature to another'' \cite{Bloom_01}. The    parameters of the model  are the number of agents in the system $L$   and the copy or imitation propensities $p_a \in \left [0,1 \right ]$ of agent  $a=1, \ldots, L$.  Previous analyses  have considered the homogeneous case only, $p_a = p ~\forall a$ \cite{Fontanari_14,Fontanari_15a}. Here we focus on the case  that the copy propensities are random variables instead, and measure 
the system performance by the time $t^*$ the system requires to find the solution of the task.

We  find that endowing the agents with  different copy propensities can greatly reduce the chances that the system is temporarily trapped in sub-optimal solutions (local maxima), which is a very likely outcome of  the imitative  search for large homogeneous systems \cite{Fontanari_14,Fontanari_15a}.  However, in the regime of system sizes where that search strategy is more effective than  the independent search, diversity  impairs the system performance  and  the optimal performance is achieved by the  homogeneous system.

\section{The task}

The task posed to the agents is  to find the unique global maximum of  a fitness landscape generated using Kauffman's 
NK model  \cite{Kauffman_87}. This model allows the  tuning  of the ruggedness of the landscape -- and hence of the difficulty of the task -- by changing the integer parameters  $N$ and $K$. More pointedly,  the NK landscape is defined in the space of binary strings of length $N$ and so this parameter determines the size of the solution space, namely, $2^N$.  The other parameter  $K =0, \ldots, N-1$ is the degree of epistasis that  has a direct    influence on the number of local maxima on the landscape. In particular,
for $K=0$ the (smooth) landscape has a single maximum, whereas for $K=N-1$, the (uncorrelated) landscape  has on the average  $2^N/\left ( N + 1 \right)$ maxima
with respect to single bit flips and the NK model reduces to the Random Energy model \cite{Derrida_81}.

The NK model associates a fitness value $\Phi \left ( \mathbf{x}  \right ) $ to each binary string $\mathbf{x} = \left ( x_1, x_2, \ldots,x_N \right )$,  with $x_i = 0,1$, which  is given by an average  of the contributions from each entry of the string, i.e.,
\begin{equation}
\Phi \left ( \mathbf{x}  \right ) = \frac{1}{N} \sum_{i=1}^N \phi_i \left (  \mathbf{x}  \right ) ,
\end{equation}
where $ \phi_i$ is the contribution of entry $i$ to the  fitness of the string $ \mathbf{x} $. The quantity $ \phi_i$ depends on the state $x_i$  as well as on the states of the $K$ right neighbors of $i$, i.e., $\phi_i = \phi_i \left ( x_i, x_{i+1}, \ldots, x_{i+K} \right )$ with the arithmetic in the subscripts done modulo $N$. This is the reason the parameter $K$ is known as the degree of epistasis: it measures the degree of interaction (epistasis) among entries of the strings. In addition,   we assign to each $ \phi_i$ a uniformly distributed random number  in the unit interval \cite{Kauffman_87}, which guarantees  that $\Phi \in \left ( 0, 1 \right )$ has a unique global maximum. Finding this maximum for $K>0$ is 
a NP-complete problem \cite{Solow_00}, which  means that the time required to solve the problem using any currently known deterministic algorithm increases exponentially fast with the length $N$ of the strings \cite{Garey_79}. For $K=0$ the sole maximum of $\Phi$ is easily located by picking for each entry $i$ the state $x_i = 0$ if  $\phi_i \left ( 0 \right ) >  \phi_i \left ( 1 \right )$ or the state  $x_i = 1$, otherwise. Finally, we note that the correlation between the fitness of any two neighboring configurations (i.e., configurations that differ by a single entry) 
is $\mbox{corr} \left ( \Phi \left ( \mathbf{x}^q  \right ),   \Phi \left ( \mathbf{x}^b  \right ) \right ) = 1 - \left ( K+1 \right )/N $, where
$\mathbf{x}^a = \left ( x_1, \ldots,x_i, \ldots x_N \right )$ and $\mathbf{x}^b = \left ( x_1, \ldots, 1-x_i, \ldots, x_N \right )$, regardless of the
value of the entry $i$.

\section{The Agents}
 
We consider a  system  composed of $L$ agents and assume that each agent can interact with all the others
(see \cite{Fontanari_15b} for the study of  more complex connection patterns).    Each agent operates in an initial binary string drawn at  random with equal probability for the bits $0$ and $1$.  At any trial $t$, agent $a $ can choose  between two distinct processes to operate  on its associated string. 

The first process,
which happens with probability $1-p_a$,  is  a random, exploratory  move in the solution space that consists of flipping a single  randomly selected bit of the binary string. The second process, which  happens with probability $p_a$, is the  imitation  of a model string, i.e.,  the string with the highest fitness value  among the  $L$  strings at  the trial. The imitation or copy process is implemented as follows.
First the target string $a$   is compared with the model string $m$  and the different bits are singled out. Then the agent  selects at random one of the distinct bits and flips it so that this bit is now the same in both  strings. As a result of the imitation process the target string becomes more similar to the model string. In the case the string $a$ is identical to the model string,  the agent executes the exploratory move with probability one. 

The imitation procedure  was motivated by the mechanism used to simulate the influence of an external media  \cite{Shibanai _01,Peres_11} in  the celebrated agent-based model proposed by  Axelrod to study the process of culture dissemination  \cite{Axelrod_97}. This procedure sets our model apart from a  similar model studied in the  Management and Organizations literature \cite{Lazer_07} (see also \cite{Herrmann_14}) where the imitation mechanism is such that  the target string becomes identical to the model string after imitation. This non-incremental change  may  permanently stuck the search in a  local maximum.  In that context, the exploratory move is called exploration, since the agent may generate new information, and the copying process, exploitation, since the agent uses information that is already present in the system \cite{March_91}.

The parameter $p_a \in \left [0,1 \right ]$ is the imitation or copy propensity of agent $a$. If  $p_a=0$  then agent $a$ will explore the solution space independently of the other agents. In previous  studies of this model \cite{Fontanari_14,Fontanari_15a,Fontanari_15b} we assumed  that
the $L$ agents exhibited the same imitation behavior, i.e., $p_a = p$ for $a=1, \ldots,L$.  Here we introduce variety in the behavior of the agents by
endowing them with different copy propensities. In particular, we consider the case that the $p_a$'s  are identically distributed independent random variables drawn from  the uniform probability distribution $Q_U \left ( p_a \right ) = 1$ for $p_a \in \left [ 0,1 \right ]$ and $Q_U \left ( p_a \right ) = 0$ otherwise, as well as the case that they are drawn from    the trimodal distribution 
\begin{equation}\label{trimodal}
Q_T \left ( p_a \right ) = \frac{1}{3} \delta \left ( p_a \right ) + \frac{1}{3}  \delta \left ( p_a - 1/2 \right ) + \frac{1}{3}  \delta \left ( p_a -1\right ). 
\end{equation}
Since in both cases we have $\langle p_a \rangle = 1/2$,  a suitable  homogeneous system  that can  serve as standard for gauging the
benefits of diversity  is that for which $p_a = 1/2$ for $a=1, \ldots,L$. We note that in the case of the trimodal distribution we consider only realizations which exhibit  all the three different  values of the  copy propensities, regardless of their proportions.

The search ends only when one of the agents finds the global maximum and we denote by $t^*$ the number of  trials made by the agent that found the solution.  Since the trial number $t$ is  incremented by one unit when the $L$ agents have executed  one of the two operations on its associated string,
$t^*$ stands also for the number of trials made by any one of the $L$ agents in the system.  Hence the total number of agent updates necessary to find the global maximum is  $Lt^*$ and so the  computational cost of the search can be defined as $C \equiv L t^*/2^N$, where for convenience we have rescaled $t^*$ by the size of the solution space $2^N$. We note that the update of the $L$ agents in a trial is sequential and so the model strings may change several times within the same trial.

\section{The Simulations}

Since the complexity of the task  is a key element to be considered when determining the organization that maximizes the problem-solving performance of the system
\cite{Lazer_07,Herrmann_14}, we study imitative  searches on easy tasks, i.e., smooth NK  landscapes with a single maximum ($K=0$), and on more difficult tasks, i.e.,  rugged NK landscapes  characterized by the parameters $\left (N=12,K=3 \right )$ and $ \left (N=18,K=5 \right )$.  For each realization of a fitness landscape we carry out $10^5$ searches starting from different initial conditions
(initial strings and copy-propensity realizations)  and the averaged results are then averaged again over at least $10^3$  realizations of landscapes characterized by the same values of the parameters  $N$ and $K>0$.  In the case $K=0$ all landscapes are equivalent and so we do not need to average over different realizations. 

\subsection{Smooth Landscapes}

\begin{figure}[!ht]
  \begin{center}
\includegraphics[width=0.48\textwidth]{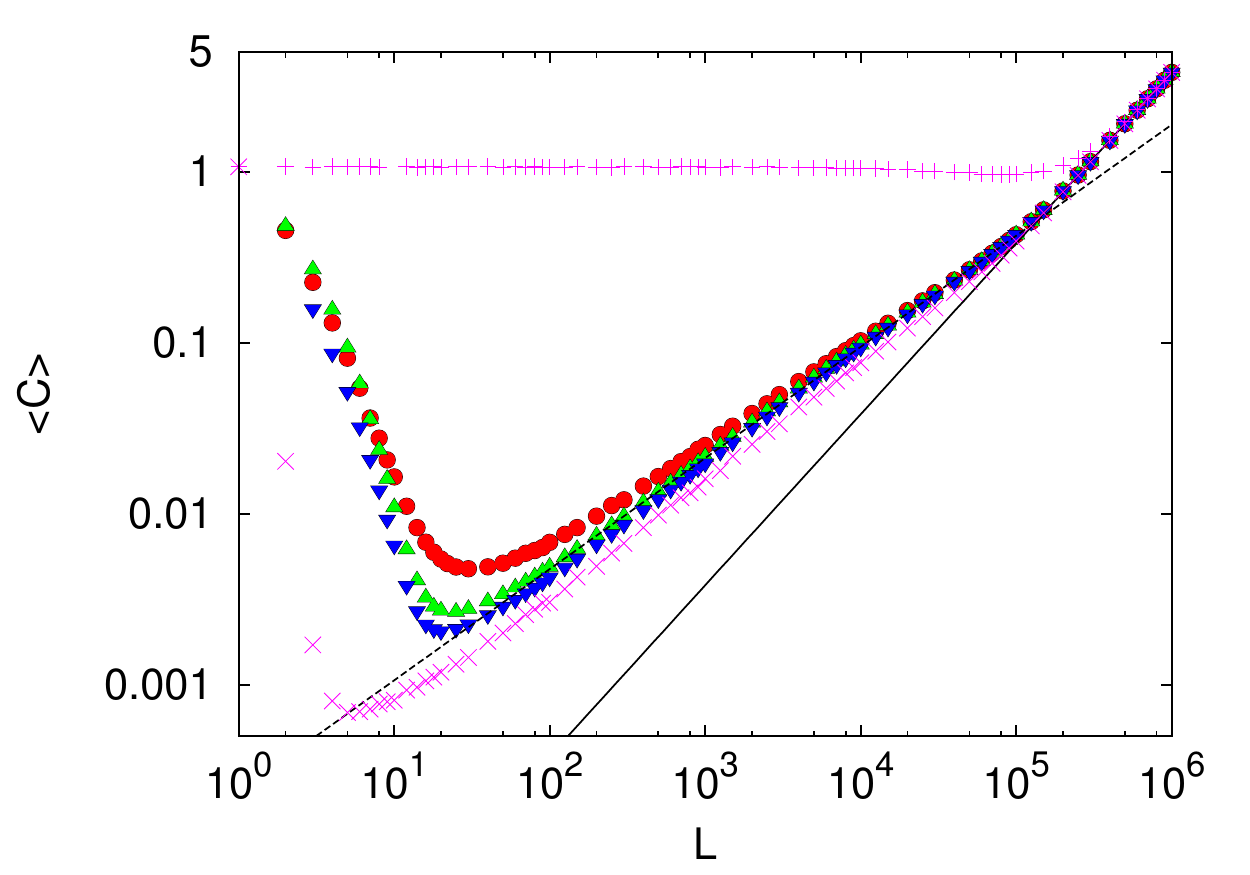}
  \end{center}
\caption{(Color online) 
Mean  computational cost $\langle C \rangle$  as function of the system size $L$ for a system of identical 
agents with $p_a=0.5 ~\forall a $  ($\CIRCLE$),  a system of agents with $p_a$ uniformly distributed in the unit interval
($\blacktriangle$), and a system of agents with $p_a$ generated using a trimodal distribution  ($\blacktriangledown$). The symbols $(+)$
are the results for the independent search ($p_a=0 ~\forall a$) and the symbols $(\times)$ for the imitative search with $p_a=1 ~\forall a$.
The solid line is the linear function $\langle C \rangle = L/2^{18}$, and  the dashed line 
 is the fitting $\langle C \rangle = 0.79 \left (L/2^{18} \right )^{0.65}$  of the uniform distribution data in the range $L \in \left [10^3,10^5 \right ]$.
 The parameters of the smooth NK landscape are
$N=18$ and $K=0$.
 }
\label{fig1_M18_K0}
\end{figure}


In Fig.\ \ref{fig1_M18_K0} we show the performances, as measured by  the mean computational cost $\langle C \rangle$, 
of  a system composed of identical agents (i.e., $p_a = 0.5$ for all agents), a system composed of agents with $p_a$ drawn from the uniform distribution $Q_U \left ( p_a \right) $
and  a system of agents with $p_a$  drawn from the (biased)  trimodal distribution $Q_T \left ( p_a \right) $. As observed in previous analyses of the imitative search \cite{Fontanari_14,Fontanari_15a,Fontanari_15b}, for each condition  there is a system size    at which  the computational cost
is minimum. We note that, for a landscape without local maxima, the best performance of the imitative search is achieved by setting $p_a =1$ for all agents (see  Fig.\ \ref{fig1_M18_K0}), since copying the fittest string at the trial is always a certain step towards the solution of the problem  \cite{Fontanari_15a}.  This is the reason  
the trimodal distribution gives the best performance among the three distributions with $\langle p_a \rangle = 0.5$ exhibited in  Fig.\ \ref{fig1_M18_K0}: it simply produces systems with a  large proportion of
 experts (i.e., agents with $p_a =1$). In fact, we have verified that a bimodal distribution, in which half of the agents have $p_a =0$ and the other half $p_a =1$, yields a better performance than the trimodal distribution.

For $L$ greater than the optimal system size, we observe two distinct growth regimes of the computational cost. The first regime, which occurs for $L < 2^N$ and  holds over for nearly three decades for the data of Fig.\ \ref{fig1_M18_K0}, is characterized by a sublinear
growth   $\langle C \rangle \sim L^{\alpha} $ with $\alpha < 1$ and signals a scenario of mild negative synergy among the agents since the time $t^*$ necessary to find the global maximum decreases
with  $L ^{\alpha-1}$ rather than with $L^{-1}$ as in the case of the independent search (absence of synergy). 
Although this regime is important because for large $N$ it is the only
growth regime that can be observed in the simulations,  the specific value of the exponent $\alpha$  is not very
informative since it depends on the distribution of the copy propensities (see Fig.\ \ref{fig1_M18_K0}) and it  increases with increasing $N$. For instance, for the uniform distribution we found $\alpha \approx 0.51$ for $N=12$,
$\alpha \approx 0.65$ for $N=18$ and  $\alpha \approx 0.72$ for $N=22$. 
The second regime, which takes place
for $L  > 2^N$,  is described by the linear function $\langle C \rangle = L/2^N$ and corresponds to the situation where the system size is so large that the solution is 
found in the first trials. In this regime, $t^*$  is not affected by the value of $L$, i.e., adding more agents to the system does not decrease
the time required to find the solution. Finally, we note that for $K=0$ the imitative  search  always performs better than the independent search. 

\begin{figure}[!ht]
  \begin{center}
\includegraphics[width=0.48\textwidth]{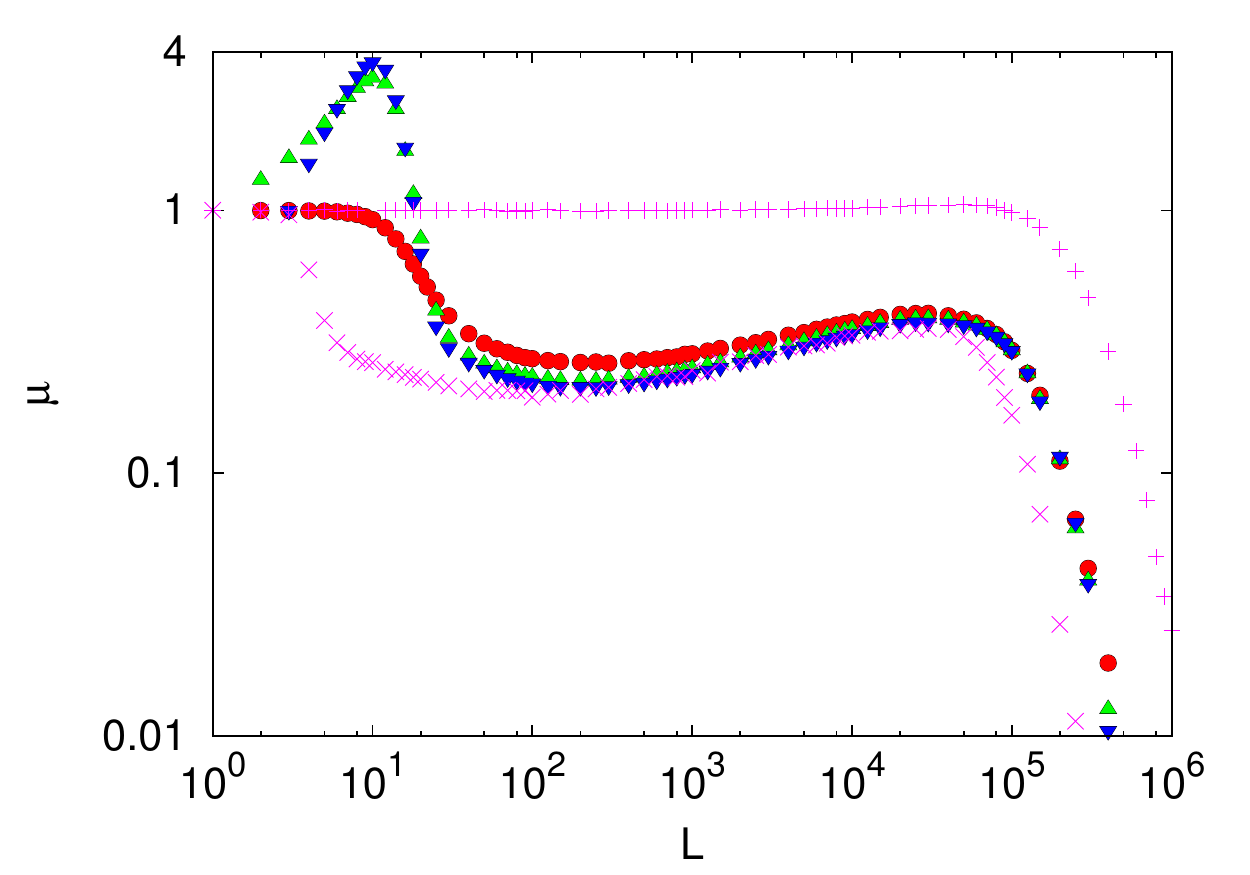}
  \end{center}
\caption{(Color online) Ratio $\mu$   between the standard deviation of the computational cost and its mean value 
 as function of the system size $L$ for the  different system compositions shown in Fig.\ \ref{fig1_M18_K0}.  The symbols convention and the
 parameters of the  NK landscape are the same as for that figure.
 }
\label{fig5_M18_K0}
\end{figure}

It is also instructive to consider the ratio between the standard deviation of the computational cost and its mean value, i.e., $\mu = \left [ \langle C^2 \rangle/\langle C \rangle^2 -  1\right ]^{1/2}$, which is shown in Fig.\ \ref{fig5_M18_K0}. Clearly, $\mu$  also gives
the ratio between the standard deviation of the time required to find the  global maximum $t^*$ and its means value.   In the regime that $\mu  \approx 1$, the  distribution of the computational cost can be well described by an  exponential  distribution, though we note that the correct distribution in the case of the independent search is the geometric distribution \cite{Fontanari_15b}. The high dispersion observed for small  size heterogeneous systems  is due to the great dispersion of the system composition, a factor  whose effect decreases as $L$ increases. We recall that this effect is absent for $L=3$ in the case the  propensities are  generated by the trimodal distribution because we consider only realizations where the three classes are represented in the system. This effect aside, it was  expected that the dispersion of $t^*$  would be smaller than for the independent search: given the smoothness and non-degeneracy of the landscape, the searches should follow neighboring paths in the solution space.  It is interesting that although the search strategies  exhibit the same mean computational cost in the regime  $L \gg 2^N$, the imitative
search  has a much smaller dispersion than the independent search.

\begin{figure}[!ht]
  \begin{center}
\includegraphics[width=0.48\textwidth]{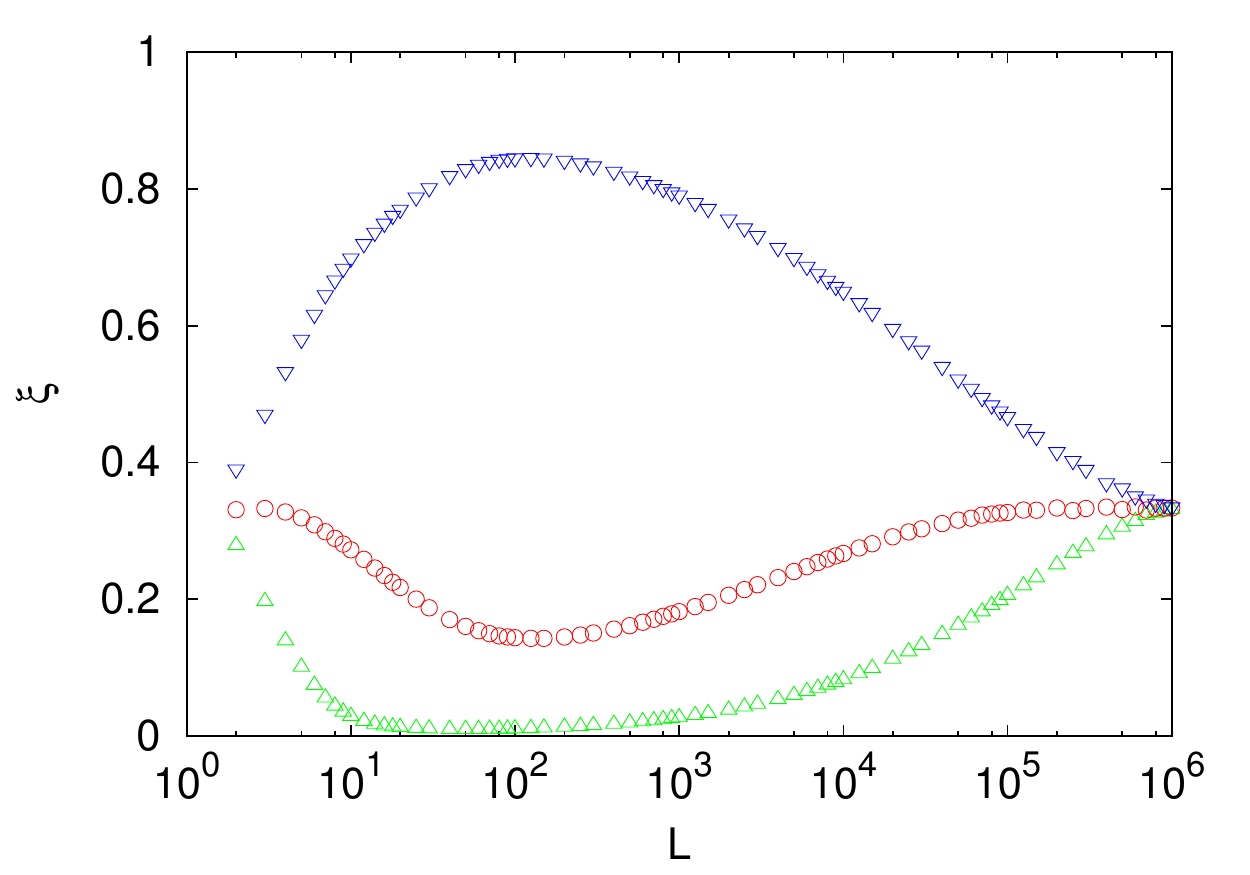}
  \end{center}
\caption{(Color online) Fraction of the searches for which the global maximum was found by agents with  low-copy ($\triangle$), average-copy ($\circ$) and
high-copy ($\triangledown$) propensities
 for the case  $p_a$ is uniformly distributed in the unit interval.
 The parameters of the smooth NK landscape are
$N=18$ and $K=0$. 
 }
\label{fig2_M18_K0}
\end{figure}

An interesting issue that we can address  in the case of heterogeneous systems is whether there is a correlation between the propensity of an agent to imitate $p_a$ and its
chances of finding the global maximum. To treat this issue  for the  case the propensities are generated by the uniform distribution $Q_U \left ( p_a \right )$
we  divide the agents in three classes, namely, low-copy propensity agents characterized by $p_a \in \left [ 0,1/3 \right )$, average-copy propensity agents for which
$p_a \in \left [ 1/3,2/3 \right )$ and high-copy  propensity agents for which $p_a \in \left [ 2/3,1 \right ]$. Figure \ref{fig2_M18_K0} shows the probability  $\xi$ that an agent belonging to one of those classes hits the global maximum. This figure corroborates our preceding remark  that  for a smooth landscape the best strategy for the agents is to copy the model string, since  that string always displays  faithful information about the location of the global maximum.  For $L > 2^N$ the determining factor  for an agent  to hit the solution  is its
proximity to the global maximum when the initial strings are set randomly and so all copy propensity classes perform equally in this regime, as expected.
The results for the trimodal  distribution $Q_T \left ( p_a \right )$
are qualitatively the same as those shown in Fig.\  \ref{fig2_M18_K0}, except that the high-copy propensity class, which in this case is characterized by $p_a =1$, has a slightly higher probability of finding the global maximum than it has for the uniform distribution.

\subsection{Rugged Landscapes}

The study of the performance of the imitative search on rugged landscapes is way more compute-intensive than on smooth landscapes for two reasons: first, the number of trials $t^*$ to hit the solution for system sizes  near the optimal size  is about 100 times greater than for smooth landscapes. Second, now we need to average the results over many (at least, $10^3$) realizations  of the NK landscape. Hence
to grasp  the behavior of the system in all  regimes  of   $L$  studied before, we  will consider first a rather small  landscape with parameters $\left (N=12,K=3 \right )$ and then verify whether the results  hold true for a larger landscape  with parameters $\left (N=18,K=5 \right )$. Note that for both landscapes the correlation between the fitness of neighboring strings is $2/3$.

\begin{figure}[!ht]
  \begin{center}
\includegraphics[width=0.48\textwidth]{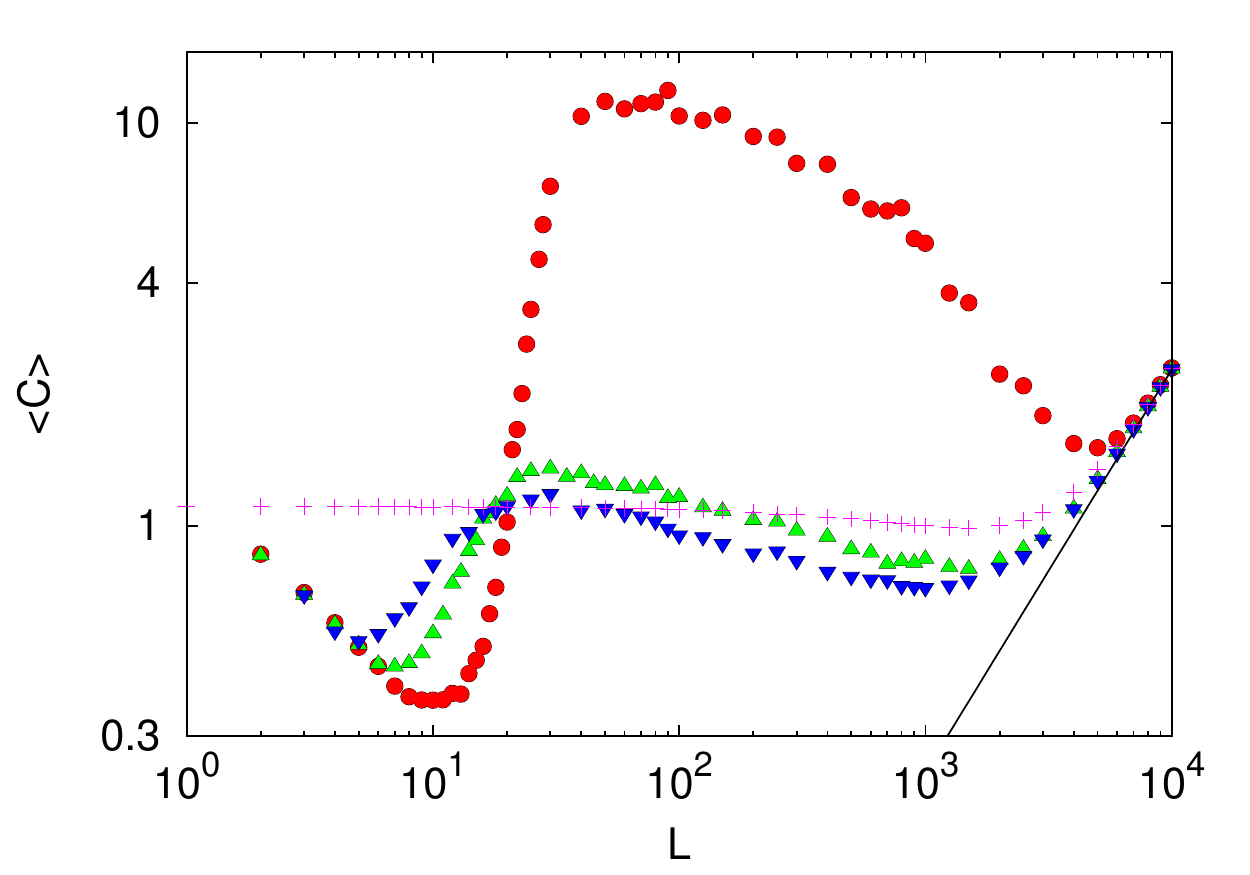}
  \end{center}
\caption{(Color online) Mean  computational cost $\langle C \rangle$  as function of the system size $L$ for a  system  of identical
agents with  $p_a=0.5~\forall a $  ($\CIRCLE$),  a system of  agents with $p_a$ uniformly distributed in the unit interval
($\blacktriangle$), and  a system of  agents with $p_a$ generated using a trimodal distribution  ($\blacktriangledown$). 
The symbols $(+)$
are the results for the independent search $p_a=0~\forall a $. The solid line is the linear function $\langle C \rangle = L/2^{12}$. 
 The parameters of the rugged NK landscape are
$N=12$ and $K=3$.
 }
\label{fig1_M12_K3}
\end{figure}

Figure \ref{fig1_M12_K3} summarizes our results for the NK landscape with parameters $\left (N=12,K=3 \right )$. This figure reveals that 
moderately large (i.e., $L \in \left [20, 2000 \right ]$)  homogeneous systems   can  easily  be trapped by the local  maxima,  from which escape can be extremely   costly.  This is akin to the groupthink phenomenon \cite{Janis_82},  when everyone in a group starts thinking alike,  which can occur when people put unlimited faith in a talented leader (the model strings, in our case). The finding that these traps can be circumvented by endowing the agents with different parameters of the behavioral rules  is a main thrust of the arguments pro diversity to boost system performance  \cite{Page_07}.  
Our results corroborate that viewpoint since  the  heterogeneous systems  exhibit an overall performance slightly  superior to the independent search, which, we recall, is not affected by the presence or absence of local maxima. The surprising finding, however, is that  for  small system sizes, where the imitative search  can be said to be efficient in the sense that $t^*$ decreases superlinearly with increasing $L$ \cite{Huberman_90,Clearwater_91}, the homogeneous system performs best.  

\begin{figure}[!ht]
  \begin{center}
\includegraphics[width=0.48\textwidth]{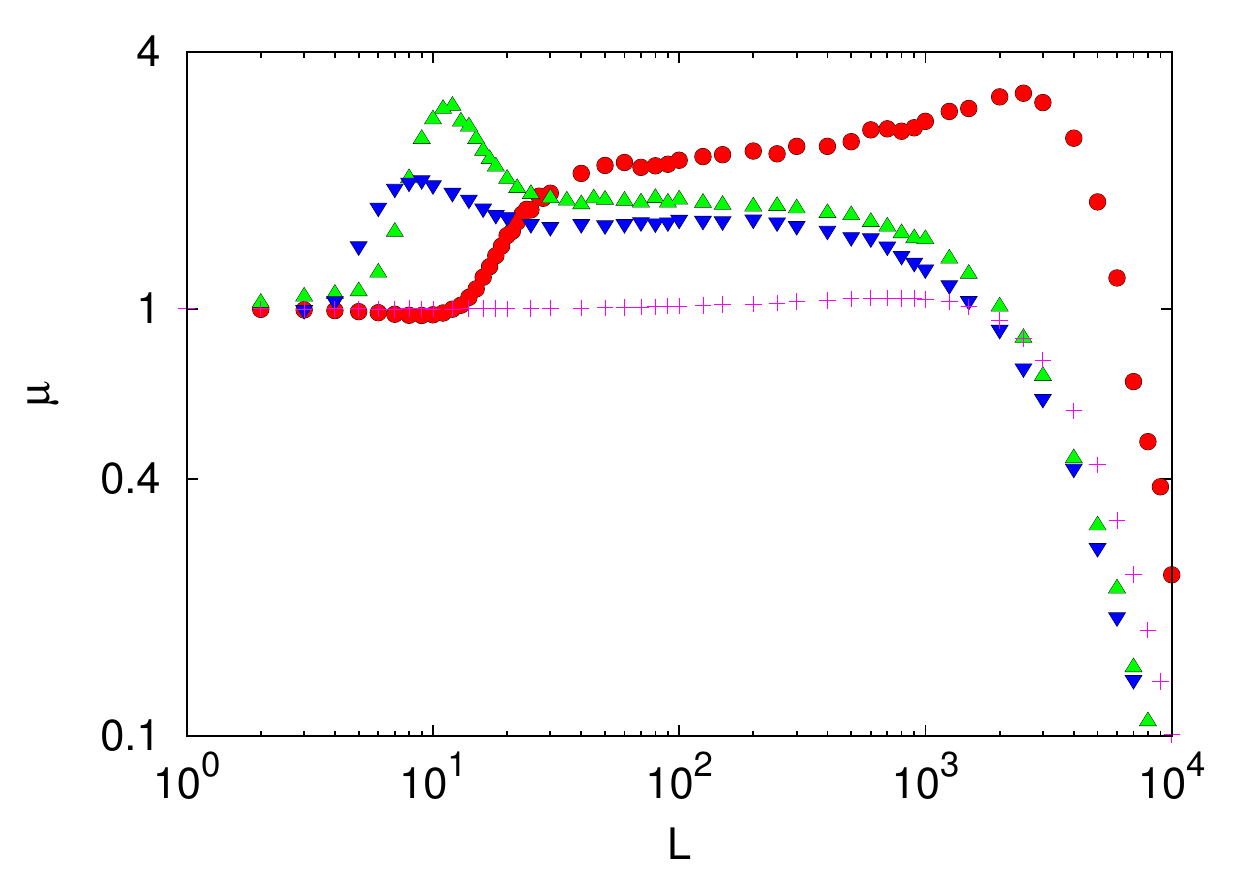}
  \end{center}
\caption{(Color online) Ratio $\mu$   between the standard deviation of the computational cost and its mean value 
 as function of the system size $L$ for the different system compositions shown in Fig.\ \ref{fig1_M12_K3}.  The symbols convention and the
 parameters of the  NK landscape are the same as for that figure.
 }
\label{fig5_M12_K3}
\end{figure}

The properly rescaled deviations around the mean values of the computational cost are shown in Fig.\ \ref{fig5_M12_K3}. As opposed to the results  for the smooth landscape (see Fig.\ \ref{fig5_M18_K0}), the deviations now are larger than those for the independent search for almost all values of $L$.  The reason is that the  searches that are (temporarily)  trapped by local maxima  contribute with very large costs, whereas the searches that avoid those traps can reach the global maximum very quickly. We note that $\mu$ does  not account for fluctuations of the computational cost due to the different landscape realizations: the standard deviation as well as the mean of the computational cost are  measured for each landscape realization and then their ratio is averaged over the different realizations.

\begin{figure}[!ht]
  \begin{center}
\includegraphics[width=0.48\textwidth]{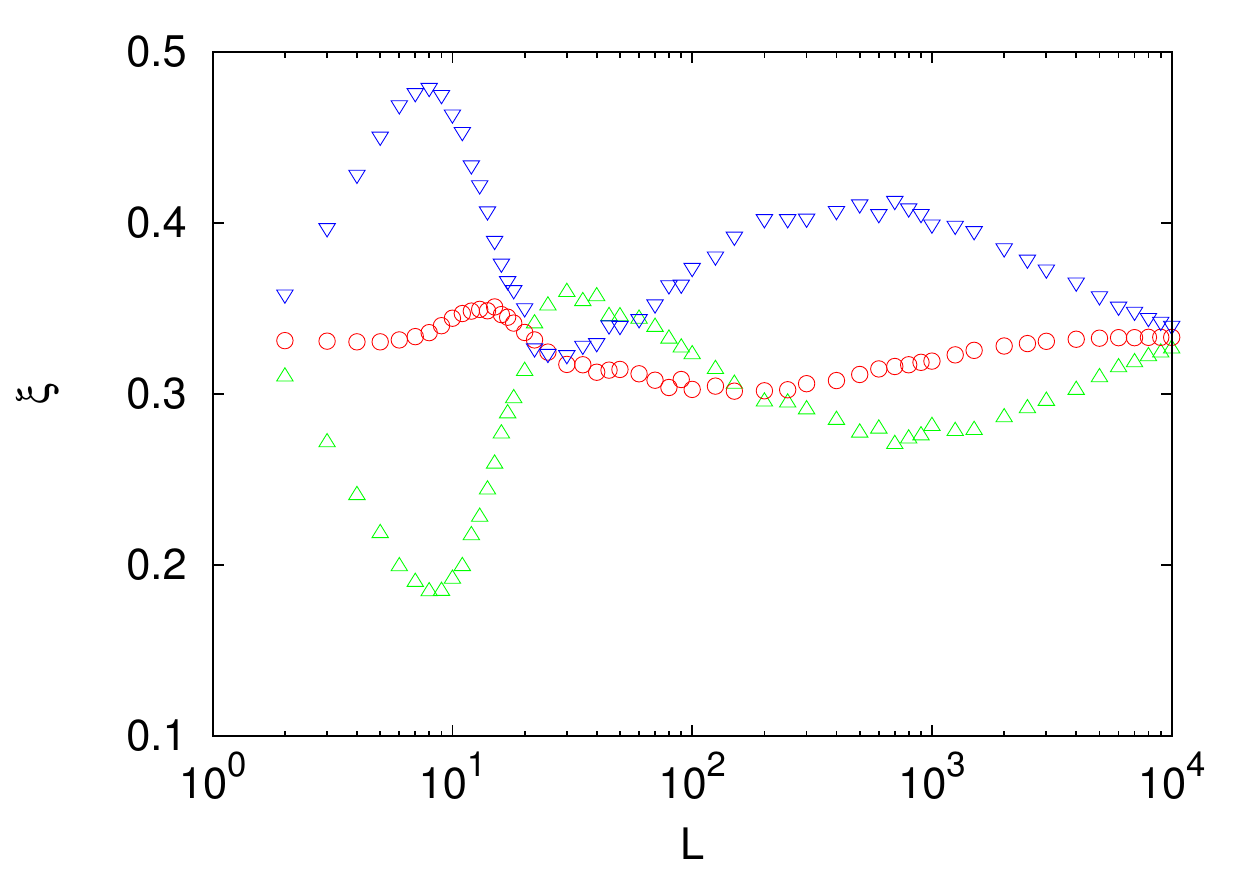}
  \end{center}
\caption{(Color online) Fraction of the searches for which the global maximum was found by agents with  low-copy ($\triangle$), average-copy ($\circ$) and high-copy ($\triangledown$) propensities
 for the case  $p_a$ is uniformly distributed in the unit interval.
 The parameters of the rugged NK landscape are
$N=12$ and $K=3$.
 }
\label{fig2_M12_K3}
\end{figure}

The chances that  agents belonging to the low-copy, average-copy or high-copy propensity  classes  hit the global maximum are shown
in Fig.\ \ref{fig2_M12_K3} in the case  $p_a$ is drawn from the uniform distribution $Q_U \left ( p_a \right )$. The situation now is way more complex than for the smooth landscape. In this case the minimum cost occurs for $L =7$, which coincides with the system size at which 
the probability that 
an agent in the  high-copy class hits the solution is maximum. Interestingly, the agents in the low-copy class are the most likely to hit the solution  in the region  where the imitative search is outperformed by the independent search, i.e., for the values of $L$ where the  system seems to be more susceptible to the presence of the local maxima. The results are qualitatively the same for the trimodal distribution.

\begin{figure}[!ht]
  \begin{center}
\includegraphics[width=0.48\textwidth]{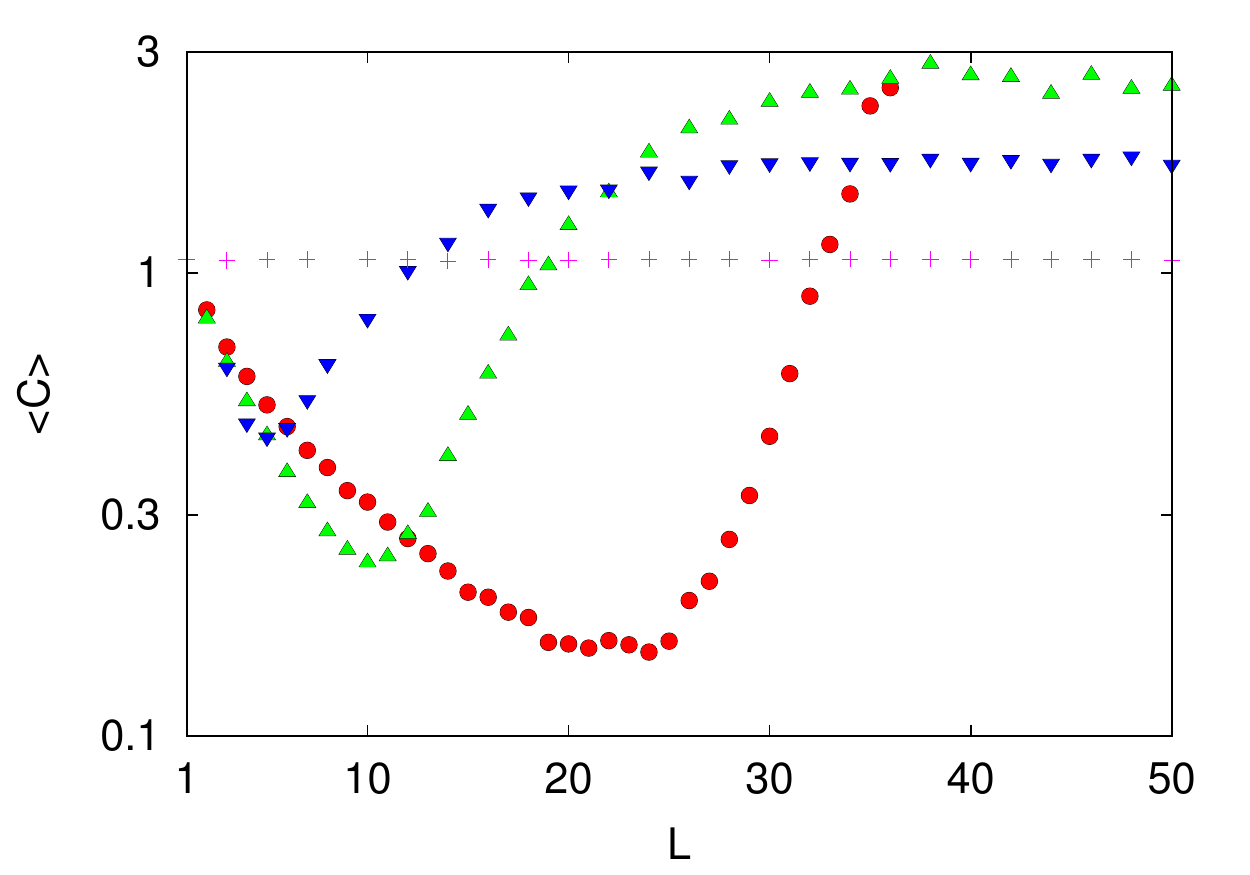}
  \end{center}
\caption{(Color online) Mean  computational cost $\langle C \rangle$  as function of the system size $L$ for a system of identical 
agents with $p_a=0.5 ~\forall a $ ($\CIRCLE$),  a system of agents with $p_a$ uniformly distributed in the unit interval
($\blacktriangle$), and a system of agents with $p_a$ generated using a trimodal distribution  ($\blacktriangledown$). The symbols $(+)$
are the results for the independent search $p_a=0~\forall a $.
 The parameters of the rugged NK landscapes are
$N=18$ and $K=5$.
 }
\label{fig1_M18_K5}
\end{figure}

The average performance of the  imitative search on  rugged landscapes characterized by the parameters  $\left (N=18,K=5 \right )$  is shown in Fig.\ \ref{fig1_M18_K5} for  small system sizes. These results are qualitatively the same as those shown in Fig.\ \ref{fig1_M12_K3},  except that a tendency that is barely visible in that figure  becomes evident now: the heterogeneous systems exhibit the best  performance for very small ($L < 10$) system sizes. Nevertheless, the advantage of the homogeneous system is striking for sizes in the range $L \in [ 10,30 ]$, corroborating the puzzling finding that if the system size can be adjusted to  maximize performance then the homogeneous system performs better than the heterogeneous one, given the constraint that $\langle p_a \rangle$ is the same in all conditions.
  The results regarding the dispersion around the mean computational cost and the  chances of agents in the different copy propensity classes to  hit the solution are qualitatively the same as those discussed for the landscapes with parameters  $\left (N=12,K=3 \right )$.
  
Finally, we note that since finding the global maxima of NK landscapes  with $K>0$ is an NP-Complete problem \cite{Solow_00}, one should not expect that the imitative  search (or any other search strategy, for that matter)   would find those maxima much more rapidly
than the independent search.

\section{Conclusion}

Our findings corroborate, in part, the prevalent views on the effects of diversity on  the efficiency of cooperative problem-solving   systems
\cite{Page_07}. In particular, in the case of easy tasks, modeled here by smooth landscapes without local maxima,  for which there is an optimal imitation strategy,  the best performance is
achieved by a homogeneous system of  agents equipped with that strategy, the so-called experts  (see Fig.\ \ref{fig1_M18_K0}). In the case
of difficult tasks, modeled by landscapes plagued of local maxima, 
we find  that diversity  is a palliative for the main deficiency of the imitative search strategy, namely, the lure of the model strings in the vicinity of the local maxima, a phenomenon analogous to Groupthink \cite{Janis_82}.  
In fact, for some system sizes diversity may produce a more than  tenfold decrease of  the computational cost  in comparison with that  of  homogeneous systems. We note, however, that  a  more efficient strategy to bypass Groupthink is to reduce the influence of the model string by decreasing the connectivity of the network \cite{Fontanari_15b}.

The main result of this paper is the surprising finding 
that if one is allowed to adust the system size $L$ to maximize the performance, then the homogeneous system will outperform
the heterogeneous ones. To offer a clue to understand this finding, we note that the optimal size of the homogeneous system is $L^* \approx N$ (see Figs.\ \ref{fig1_M18_K0}, \ref{fig1_M12_K3} and \ref{fig1_M18_K5}), which means that
the optimal system is composed of a model string together with a cloud of mutant strings that differ from it typically by one or two entries.  Since this is the manner viral quasispecies explore  their fitness landscapes \cite{Domingo_12}, it is probably the optimal (or near-optimal) way to explore rugged fitness landscapes.

\acknowledgments

This research was partially supported by grant
2015/21689-2, S\~ao Paulo Research Foundation
(FAPESP) and by grant 303979/2013-5, Conselho Nacional de Desenvolvimento 
Cient\'{\i}\-fi\-co e Tecnol\'ogico (CNPq). The research used resources of the LCCA - Laboratory of Advanced Scientific Computation of the University of S\~ao Paulo.

\end{document}